\documentclass[superscriptaddress,nobibnotes,nofootinbib,floatfix,aps,prm,citeautoscript,preprint]{revtex4-1}

\usepackage[draft]{changes}
\definechangesauthor[name={Yi}, color=blue]{Yi}

\usepackage{graphicx}
\usepackage{epsfig}
\usepackage{subfigure}
\usepackage{color}
\usepackage{latexsym}
\usepackage{amsmath,bm,multirow}
\usepackage{amsfonts}
\usepackage{dcolumn}           
\usepackage{natbib}
\usepackage[colorlinks,linkcolor=blue,citecolor=green]{hyperref}
\usepackage{physics}
\usepackage{siunitx}

\newcommand{\etal}{\textit{et al.~}}

\begin{document}
\title{Computational prediction of lattice thermal conductivity -- a comparison of molecular dynamics and Boltzmann transport approaches}

\author{Marcello Puligheddu}
\thanks{These two authors contributed equally}
\affiliation{Institute for Molecular Engineering, University of Chicago, Chicago, IL}

\author{Yi Xia}
\thanks{These two authors contributed equally}
\affiliation{Center for Nanoscale Materials, Argonne National Laboratory, Argonne IL}

\author{Maria Chan}
\email{mchan@anl.gov}
\affiliation{Center for Nanoscale Materials, Argonne National Laboratory, Argonne IL}

\author{Giulia Galli}
\email{gagalli@uchicago.edu}
\affiliation{Institute for Molecular Engineering, University of Chicago, Chicago, IL}
\affiliation{Material Science Division, Argonne National Laboratory, Argonne, IL}
\affiliation{Department of Chemistry, University of Chicago, Chicago, IL}
\date{\today}

\begin{abstract}
The predictive modeling of lattice thermal conductivity is of fundamental importance for the understanding and design of materials for a wide range of applications. Two major approaches, namely molecular dynamics (MD) simulations and calculations solving approximately the Boltzmann transport equation (BTE), have been developed to compute the lattice thermal conductivity. We present a detailed direct comparison of these two approaches, using as prototypical cases MgO and PbTe. The comparison, carried out using empirical potentials, takes into account the effects of fourth order phonon scattering, temperature-dependent phonon frequencies (phonon renormalization), and investigates the effects of quantum vs. classical statistics. We clarify  that equipartition, as opposed to Maxwell Boltzmann, govern the statistics of phonons in MD simulations.  We find that  lattice thermal conductivity values from MD and BTE show an apparent, satisfactory agreement; however such an agreement is the result of error cancellations. We also show  that the primary effect of statistics on thermal conductivity is via the scattering rate dependence on phonon populations. 
\end{abstract}

\maketitle

\section{Introduction}
Thermal transport properties are important for myriad materials, including thermoelectrics, (opto)electronics, photovoltaic and photoelectrochemical cells, and batteries. To facilitate fundamental understanding and materials design, the accurate prediction of thermal transport coefficients is critical. A multitude of methodologies have been introduced to facilitate such prediction. Of particular significance are two categories of methods,  based either on considering atomic motion (molecular dynamics, MD) or collective vibrational excitations, i.e. phonons in solids (anharmonic lattice dynamics, ALD). 

Over the years, several MD-based approaches have been developed to compute the thermal conductivity in bulk crystalline and amorphous materials, nanostructures, and fluids, with the central premise of tracking heat flow through atomic motion.
In the Green-Kubo (GK) method, based on the fluctuation-dissipation theorem, the thermal conductivity is calculated from the fluctuations of the heat correlation function during an equilibrium simulation, in the NVE ensemble.
Non Equilibrium (NE) methods compute the thermal conductivity in the steady state from the response of the system to a perturbation. They can be classified according to the nature of the perturbation: in the older direct method it is a temperature gradient causing a heat flux \cite{Evans2007}; in reverse NEMD, also known as the Muller-Plathe approach,the constant heat flux acts as the perturbation \cite{MullerPlathe1997}.
The "approach to equilibrium" methods compose the third class of MD based method. In these approaches, the thermal conductivity is computed from the time response of the system to an instantaneous perturbation. This perturbation can be in the form of a square \cite{Lampin2013, Melis2014} or sinusoidal temperature profile \cite{Puligheddu2017}.
Meanwhile, phonon-based ALD approaches for calculating lattice thermal conductivity, applicable to crystalline systems, have been developed based on the Boltzmann transport equation (BTE), with the central assumption that collective vibrations can be thought of as quasi-particles termed phonons which have explicit group velocities and scatter with each other \cite{ziman1960electrons,srivastava1990physics,peierls1996quantum}. Within these approaches, the lattice thermal conductivity is calculated by solving the BTE, under the relaxation time approximation, with relaxation times obtained using perturbation theory and considering anharmonic phonon-phonon interactions, such as three- and four-phonon scattering processes \cite{Maradudin1962,dfpt1,Deinzer2003,Broido2005,Tianli2016}.
While MD and ALD are fundamentally distinct and practically disparate approaches, they have nonetheless both been shown to give reasonably accurate predictions for similar systems \cite{Turney2009a,He2012,Tianli2016}.

Typically, GK calculations are performed using classical MD where the potential energy is given by an interatomic potential as a function of the positions of the atoms. Recently, Marcolongo et. al. \cite{Marcolongo2016} 
and Carbogno et al. \cite{Carbogno2017} showed that the Green-Kubo formalism can be applied to ab initio MD, although the method of Ref. \cite{Carbogno2017} is limited to solids and contains approximations on the partition of slow and fast vibrational modes.
We note that the Green-Kubo formalism does not rely on the definition of phonons and is generally applicable: indeed it can describe conductive heat transport in crystalline or amorphous materials, materials with defects, nanostructures, and fluids. The GK method implicitly accounts for all anharmonic terms in the potential energy. In BTE language, this would translate to including all phonons processes in the perturbative expansion (instead of only up to the third or fourth order), thus allowing for greater accuracy at high temperatures.
Nonetheless, the GK method suffers from three deficiencies: (i) being based on classical MD, it is not known how to include quantum effects such as zero point energy or quantization of the phonon energy levels, leading to lower accuracy at low temperatures. This problem is of a theoretical nature and, to our knowledge, has not been solved, although semi-empirical corrections have been proposed \cite{Wang1990, Lee1991};
(ii) long simulation times (order of ns) are needed for convergence, especially for systems with high values of thermal conductivity, mostly due to noise in the long time tail of the heat flux autocorrelation function. Acceleration is possible based on the methods proposed in Ref. \cite{Ercole2017} or \cite{Carbogno2017}, although the latter one is limited to crystals and based on phonon theory; and 
(iii) the value of the thermal conductivity calculated in a GK simulation is a function of the number of atoms included in the MD cell, converging in the infinite limit to the bulk value. Currently, no theory or model exists to predict or extrapolate to the thermodynamic limit, leading to the necessity of simulating at various cell sizes with increasing numbers of atoms until satisfactory convergence is obtained. We refer to \cite{He2012} for a detailed study.

The BTE-ALD approach is more advantageous compared with MD-based method in the following respect: (i) the quantum (Bose-Einstein) statistics is utilized, thus (presumably) attaining better accuracy at low temperatures; (ii) first-principles calculations may be used to obtain third and fourth order force constants in a relatively straightforward manner for a variety of systems, to accurately model heat transport in bulk crystals with relatively low computational costs, and computation for larger systems in some cases can be parallelized \cite{dfpt1,broido,shengbte,csld}; and (iii) convergence issues with respect to system size and sampling are less severe than in MD. 
However, BTE-ALD suffers from several deficiencies: (i) the approach depends on the assumption that lattice vibrations can be treated as quasiparticles, i.e., both the anharmonic frequency shift and broadening (real and imaginary part of the phonon self-energy) are relatively small compared to the phonon frequency, and fails when quasiparticles are not well-defined, such as in the localization limit in systems with strong intrinsic disorder \cite{AllenLocalization,Seyf2017}; (ii) the approach is based on perturbation theory, which is accurate only in the perturbative limit and may fail in severely anharmonic systems or at high temperatures since the scattering processes are limited to third or at most fourth order; and (iii) it is difficult to use the phonon BTE method to model heat transport in non-crystalline systems including defects and nanostructures, and the approach cannot be applied to fluids or even solids when diffusion occurs.

In an earlier study \cite{Turney2009a}, Turney \textit{et al.} presented and compared methods in several categories: (1) quasiharmonic and anharmonic lattice dynamics calculations, (2) a combination of quasiharmonic lattice dynamics calculations and molecular dynamics simulations, and (3) Green-Kubo and direct molecular dynamics, to assess their validity. They pointed out that the lattice dynamics calculations tend to underestimate lattice thermal conductivity at above half of the Debye temperature. However, their lattice dynamics calculations excluded higher-than-third-order anharmonic phonon-phonon interactions. In addition, the impact of different statistics (quantum vs. classical) on calculated thermal conductivity was not clarified. In a follow-on study, the same authors \cite{Turney2009b} assessed different corrections to MD simulations to account for quantum statistics, using silicon and the Stillinger Weber potential. They found that these corrections failed at low temperatures due to the classical distribution of phonon modes. He et. al.\cite{He2012} compared the GK-MD and BTE approaches for the computation of thermal conductivity of Si, Ge, and Si-Ge alloys, and found that the results are consistent for the pure compounds, though alloy systems prove to be problematic for BTE. In this study, some investigations into the effects of quantum statistics were carried out via the use of classical statistics in BTE. However, phonon occupation numbers were assumed to obey the Maxwell-Boltzmann distribution, which we will show below to be incorrect. The effects of temperature-induced anharmonic phonon renormalization or higher-than-third-order interactions were again not considered. Since several of the previous studies were carried out on Si or related systems, it would also be of interest to test the results on more diverse types of systems. 

In the current work, we aim to provide a controlled, comprehensive, and systematic comparison of MD vs. ALD-BTE based approaches for the prediction of lattice thermal conductivity. In order to accomplish a meaningful comparison, we select two representative systems: a small-gap semiconductor with low lattice thermal conductivity (PbTe), and an insulator with higher lattice thermal conductivity (MgO).  These two materials have the further advantage of having simple structures. For each material, we use MD and BTE-ALD, with the same interatomic potential, to evaluate the temperature-dependent values of the lattice thermal conductivity well below, near, and well above their respective Debye temperatures. We further investigate the source of the differences between the two approaches by deriving lifetimes from MD simulations, enumerating heat capacities and phonon occupancies under different treatment of statistics, and using these quantities in the BTE expression. The goal is to determine the underlying physical reasons for divergences and convergences between the two approaches.

\section{Methods}
\subsection{Model Systems and Interatomic Potentials}

To ensure that any differences in computed thermal conductivity arise solely from the treatment of heat transport (i.e. atoms vs. phonons), we used the same interatomic potential for both MD and BTE calculations. For MgO, we used the potential described in Ref. \cite{Shukla2008}; for PbTe, we used the potential from Ref. \cite{Qiu2012}. Both potentials are of the Buckingham-Coulomb type. The MgO potential was shown \cite{Shukla2008} to predict experimental lattice constant, thermal expansion and thermal conductivity reasonably well in the 300K-1500K range. The PbTe potential developed by Qiu~\textit{et al.}\cite{Qiu2012} was shown to reasonably reproduce the mechanical and phonon properties of PbTe bulk crystal, as well as lattice thermal conductivity. We studied MgO at 500K, 750K, and 1000K and PbTe at 100K, 150K, and 300K. The experimental Debye temperatures of MgO and PbTe are 743K  \cite{Beg:a12841} and 177K  \cite{Houston1968}, respectively. To account for thermal expansion, the lattice parameters computed from NPT MD simulations at each temperature were used. 

\subsection{Calculation of Thermal Conductivity using  Molecular Dynamics}
In the GK formalism, the thermal conductivity is given by
\begin{equation}
    \kappa = 
        \frac{1}{3V k_B T^2} 
        \int_0^\infty \langle J(0) \cdot J(t) \rangle \text{d}t
\end{equation}
where the brackets $\langle \cdot \rangle$ indicate the thermodynamic average and $J$ is the heat flux due to spontaneous fluctuations.

All classical simulations were performed using LAMMPS \cite{Plimpton1995}. The MgO samples contain 32768 (500-750K) or 4096 (1000K) atoms, with the smaller number of atoms for higher temperature due to the shorter mean free path at higher temperature. The PbTe samples contain 8192 atoms. For all systems, Nose-Hoover NVT equilibration runs of 20 ps for MgO and 100 ps for PbTe were followed by NVE simulations of 3 ns to obtain the lattice thermal conductivity ($\kappa$). For each temperature and material, 4-9 GK MD runs were performed, to give a total of 12-27 ns of statistics. The time step used for MgO is 1 fs whereas that for PbTe is 0.5 fs. The MgO 500K and 1000K data are from the Supporting Information in Ref.\cite{Puligheddu2017}.

\subsubsection{Calculation of Phonon Lifetimes}

Before starting our simulations, we compute the phonon frequencies and eigenvectors for the target temperature using the phonon renormalization (PRN) scheme described in Ref. \cite{YXPbTe2018}. Once the phonon frequencies $\omega_{\lambda}$ and eigenvectors $e_{\lambda}$ are known, the energy of each phonon mode can be computed during a MD simulation as a function of time. The energy of a phonon mode $\lambda$ is calculated as
\begin{equation}
    E_{\lambda}(t) = K_{\lambda}(t) + U_{\lambda}(t) 
\end{equation}
where $K$ and $V$ are, respectively, the kinetic and potential components of the energy of the phonon mode. The kinetic component is calculated from the projection $\dot{q}$ of the phonon mode on the velocities $v$ of the atoms. The potential component is calculated by projecting onto the displacement $u$ of the atoms from their equilibrium positions $r_0$, i.e.
\begin{equation}
    K_{\lambda}(t) = \frac{1}{2N} \| \dot{q}_{\lambda} \|^2 \qquad
    U_{\lambda}(t) = \frac{1}{2N} \omega_{\lambda}^2 \| q_{\lambda} \|^2
    \label{eq:UK}
\end{equation}
\begin{equation}
    \dot{q}_{\lambda} = \sum_{l,b} \sqrt{ \frac{m_b}{N} } e^{i \vec{k} r_0(l,b)} \vec{e}_{b,\lambda}^{\;*} \cdot v(l,b,t)
\end{equation}
\begin{equation}
          q_{\lambda} = \sum_{l,b} \sqrt{ \frac{m_b}{N} } e^{i \vec{k} r_0(l,b)} \vec{e}_{b,\lambda}^{\;*} \cdot u(l,b,t)
\end{equation}
Here, $l$ and $b$ are indices over the primitive cells and the atoms inside a primitive cell, respectively. $N$ is the number of atoms in the system, $m_b$ is the mass of atom $b$, and $\omega_\lambda$ is the frequency of the phonon mode $\lambda$. The displacement $u$ is given by $u(t) = r(t) - r_0(t)$. The vector $\vec{e}_{b,\lambda}$ describes the direction and phase of the displacement of atom $b$ due to the phonon with wavevector $\vec{k}$ and polarization $\lambda$.\\

The lifetime of each phonon mode is then calculated from the normalized autocorrelation of the energy:
\begin{equation}
    \tau_{\lambda} = \int_0^\infty \frac{\langle E_{\lambda}(0) E_{\lambda}(t) \rangle}{\langle E_{\lambda}(0) E_{\lambda}(0) \rangle} \text{d} t
    \label{eq:MD_tau}
\end{equation}
In order to reduce the noise due to the tail of the autocorrelation, we compute the lifetime from a fit to the equation 
\begin{equation}
    \frac{\langle E_{\lambda}(0) E_{\lambda}(t) \rangle}{\langle E_{\lambda}(0) E_{\lambda}(0) \rangle} = e^{-t/\tau_{\lambda}}
\end{equation}
with the lifetime $\tau_{\lambda}$ as a single fitting parameter. 
Lifetimes are computed as described above during an NVE simulation, once  the system is prepared at the target temperature using a Nose-Hoover thermostat. 
The approach described to obtain lifetimes relies on the definition of phonons in the quasi-harmonic approximation. 

\subsubsection{Average Phonon Occupation}
The classical Hamiltonian of the system can be rewritten, in the harmonic approximation, as a sum of kinetic and potential energy over phonon modes:
\begin{equation}
    H_{harm}(\{q,\dot{q}\}) = \sum_{\lambda} \big[ \frac{1}{2N} \| \dot{q}_{\lambda} \|^2 +  \frac{1}{2N} \omega_\lambda^2 \| q_{\lambda} \|^2 \big]
\end{equation}
 As a consequence of the equipartition theorem, the average kinetic and potential energy for each phonon mode is equal to $k_B T/2$. Each phonon mode has an average energy and phonon number given by:
\begin{equation}
\langle E_{\lambda} \rangle = k_B T \qquad \langle n_{\lambda} \rangle = \frac{k_B T}{\hbar \omega_{\lambda}}
\end{equation}
From the procedure used to calculate the phonon lifetimes, described above, we can obtain the phonon energy averaged over time
\begin{equation}
\langle E_{\lambda} \rangle = \frac{1}{\Gamma} \int_0^{\Gamma} E_{\lambda}(t) \text{d} t
\end{equation}
where $\Gamma$ is the total simulation time, and  the mode-by-mode average phonon occupation number $\langle n_\lambda \rangle$ is
\begin{equation}
\langle n_{\lambda} \rangle = \frac{\langle E_{\lambda} \rangle}{\hbar \omega_{\lambda}}
\label{eq:phonon_MD_n}
\end{equation}

\begin{figure}[htp]
	\includegraphics[width = 1.0\linewidth]{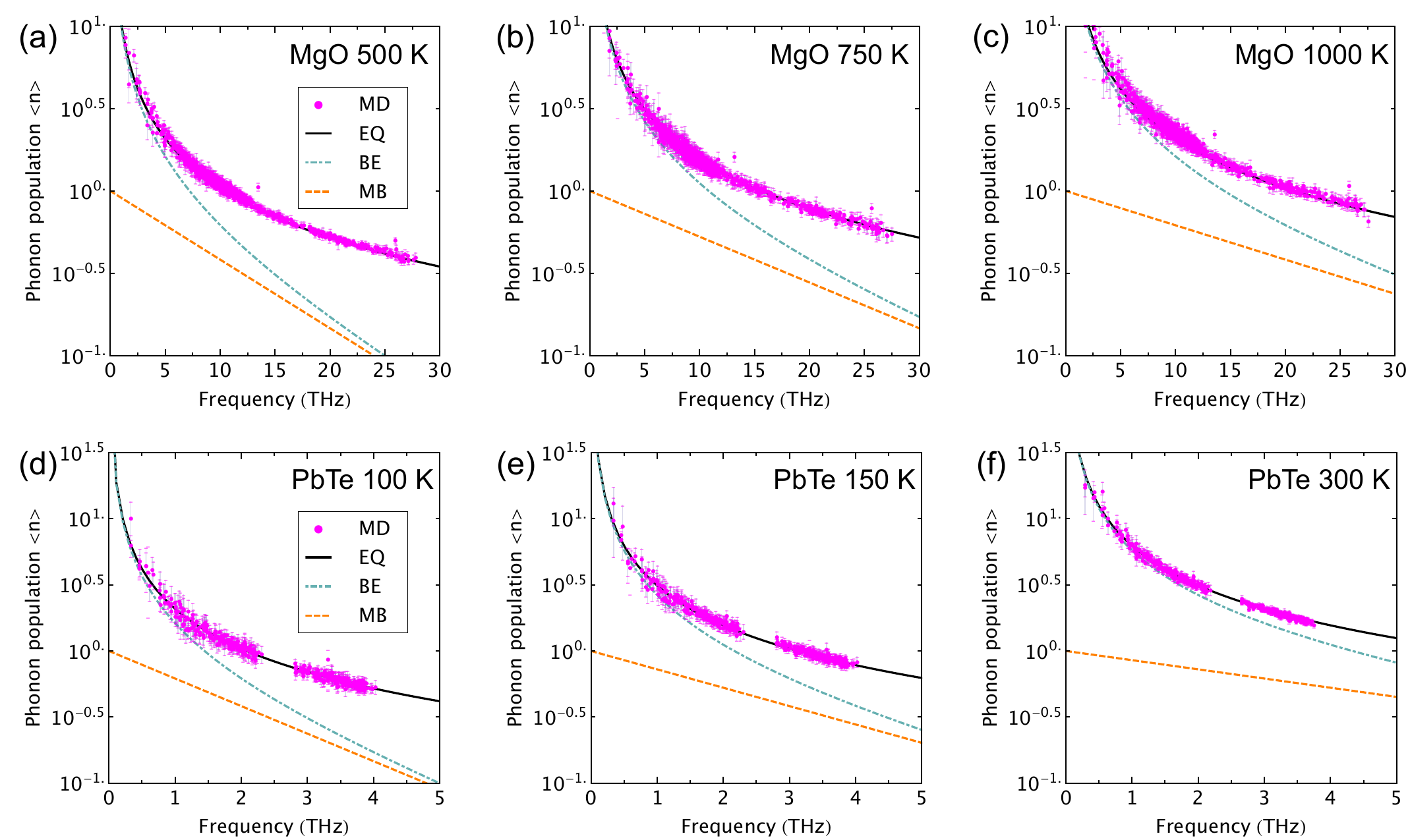}
	\caption{Phonon occupation per mode as a function of frequency for MgO at 500K (a), 750K (b) and 1000K (c) and PbTe at 100K (d), 150K (e) and 300K (f). For all materials and temperature, we compare the energy distribution calculated in our molecular dynamics (MD) simulations against three possible statistics: Bose-Einstein (BE) (blue dot-dashed lines), equipartition (EQ) (solid black lines) and Maxwell-Boltzmann (MB) (orange dashed lines).}
	\label{fig:MgO_PbTe_pop}
\end{figure}

Our results for phonon numbers  for MgO and PbTe at various temperatures are shown in Figure \ref{fig:MgO_PbTe_pop}, where we compare the distributions obtained in our calculations against those predicted by equipartition, as well as the Maxwell-Boltzmann (MB) and Bose-Einstein (BE) statistics. The computed distributions deviate significantly from those of the  MB distribution, illustrating that classical phonons in MD are distributed not via MB but equipartition (EQ). The computed distributions deviate also from BE statistics but the agreement improves with increasing temperature, as expected. 

\subsubsection{Frequency Shifts}
Phonon frequencies are dependent on temperature due to different degrees of atomic displacements from their equilibrium positions as a function of T, amounting to sampling different portions of the anharmonic potential energy surfaces. This change in phonon frequencies as a function of T (which we call shift in phonon frequencies relative to their T=0 value, or phonon renormalization) manifests in MD simulations. In our analysis, we compute the frequency shift from the kinetic part $K_{\lambda}(t)$ of the phonon mode energy. We fit its normalized time autocorrelation function to the equation
\begin{equation}
    \frac{\langle K_{\lambda}(0) K_{\lambda}(t) \rangle}{\langle K_{\lambda}(0) K_{\lambda}(0) \rangle} = e^{-t/\tau_{\lambda}} \cos^2{\omega_{\lambda}(T)}
    \label{eq:lifetime_cos_exp}
\end{equation}
where we added the dependence $(T)$ to the frequency $\omega_\lambda$. The frequency shift is then defined as the difference between the frequency obtained during MD simulations and  that obtained from diagonalizing the dynamical matrix at T=0. Note that this procedure does not include effects arising from the temperature-dependent change in lattice constant, as we compute both frequencies at the same lattice constant.

\subsection{Anharmonic Lattice Dynamics and Boltzmann Transport Equation}

\subsubsection{Boltzmann Transport Equation}
The Boltzmann transport equation (BTE) can be used to describe the time evolution of the positions and momenta of a system of particles, e.g. phonons. According to the BTE, at equilibrium the evolution of the occupation probability $n_{\lambda}$ of a specific phonon mode $\lambda$ due to diffusion, scattering, and the presence of an external heat current must balance:
\begin{equation}
\frac{\partial n_{\lambda}}{\partial t} (\mathbf{r})= \frac{\partial n_{\lambda}}{\partial t} (\mathbf{r})_{\text{diff}} + \frac{\partial n_{\lambda}}{\partial t} (\mathbf{r})_{\text{scatt}} + \frac{\partial n_{\lambda}}{\partial t} (\mathbf{r})_{\text{ext}}  = 0.
\end{equation}
Under the relaxation time approximation (RTA), the scattering term can be expressed as
\begin{equation}
-\frac{\partial n_{\lambda}}{\partial t} (\mathbf{r})_{\text{scatt}} = \frac{n_{\lambda}-n_{\lambda}^{0}}{\tau_{\lambda}},
\end{equation}
where $n_{\lambda}^{0}$ and $\tau_{\lambda}$ denote the equilibrium occupation probability and relaxation time, respectively. Under an external temperature gradient $\Delta T$, the deviation of the occupation number from its equilibrium value  $\lambda$ is given by
\begin{equation}
n_{\lambda}-n_{\lambda}^{0} = \mathbf{v}_{\lambda}\frac{\partial n_{\lambda}}{\partial T} \nabla T \tau_{\lambda},
\end{equation}
where $\mathbf{v}_{\lambda}$ is the phonon group velocity vector. 
In the linear regime, the lattice thermal conductivity tensor is defined by
\begin{equation}
    J^{\alpha}=-\sum_{\beta}\kappa_{l}^{\alpha\beta}(\nabla T)^{\beta},
\end{equation}
where $J^{\alpha}$ is the heat current along $\alpha$ direction and can be obtained as 
\begin{equation}
    J^{\alpha}=\sum_{\lambda}\int n_{\lambda}\hbar \omega_{\lambda} \mathbf{v}_{\lambda}^{\alpha} \frac{d\mathbf{k}}{(2\pi)^3},
\end{equation}
where $\mathbf{k}$ denotes the phonon wave vector. The resulting lattice thermal conductivity tensor within the RTA is
\begin{equation}
\label{eq:BTE}
\kappa_{l}^{\alpha\beta} = \frac{1}{ NV k_{\text{B}}T^2 } \sum_{\lambda} n_{\lambda}^{0} (n_{\lambda}^{0}+1) (\hbar \omega_{\lambda})^2 v_{\lambda}^{\alpha}v_{\lambda}^{\beta} \tau_{\lambda},
\end{equation}	
where $k_{\text{B}}$ is the Boltzmann constant,  $V$ is the volume of unit cell, $N$ is the total number of phonon wave vectors included in the summation, $\omega_{\lambda}$ and $v_{\lambda}$ are the frequency and group velocity of phonon mode $\lambda$, with Cartesian coordinates indexed by $\alpha$ and $\beta$. Typically, $\omega_{\lambda}$ and $v_{\lambda}$ are extracted from phonon dispersions in the harmonic approximation, assuming small atomic  displacements, by computing second-derivatives of the potential energy with respect to atomic displacements. Within the framework of anharmonic lattice dynamics, $\tau_{\lambda}$ is assumed to arise primarily from intrinsic phonon-phonon scattering events \cite{mahan2000many}, with the  lowest-order contribution being three-phonon processes.

\subsubsection{Anharmonic Lattice Dynamics} \label{34scattering}
\begin{figure}[htp]
    \centering
	\includegraphics[width = 0.6\linewidth]{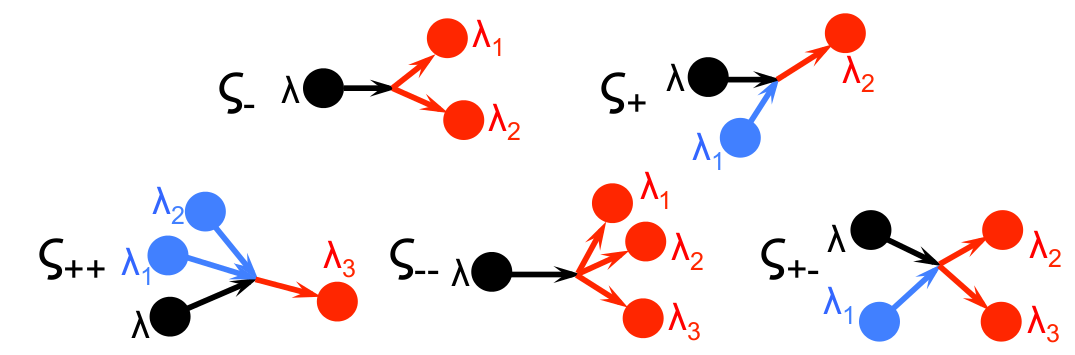}
	\caption{
	A schematic representation of three-phonon processes: (top) the splitting process ($ \zeta_{-}: \lambda \rightarrow \lambda_1 + \lambda_2 $) and the combination process ($ \zeta_{+}: \lambda + \lambda_1 \rightarrow \lambda_2 $);  and (bottom) four-phonon processes:  combination ($ \zeta_{++}: \lambda + \lambda_1 + \lambda_2 \rightarrow  \lambda_3 $), redistribution ($ \zeta_{+-} \lambda + \lambda_1 \rightarrow \lambda_2+ \lambda_3 $), and splitting ($ \zeta_{--}: \lambda \rightarrow \lambda_1 + \lambda_2+ \lambda_3 $).
	}.
	\label{fig:PhononProcess}
\end{figure}

Most recently, Feng and Ruan \cite{Tianli2016,PhysRevB.97.079901,Tianli2017} performed rigorous calculations of four-phonon scattering rates in the ALD-BTE framework including  fourth order IFCs by extending the derivation of Maradudin \textit{et al.} \cite{Maradudin1962}. Their studies reveal that even for diamond, silicon, and germanium, which are generally considered low-anharmonicity materials, the contribution of four-phonon scattering rates is comparable to that of three-phonon scattering rates near/above the Debye temperature.
Following their derivation based on Fermi's golden rule (FGR) \cite{Tianli2016}, the scattering rates ($\tau_{3,\lambda}^{-1}$ and $\tau_{4,\lambda}^{-1}$) associated with three- and four-phonon processes (see Fig.\ \ref{fig:PhononProcess}) in the single mode relaxation time approximation (SMRTA) are given by

\begin{equation}
\label{eq:ThreeP}
\begin{split}
\tau_{3,\lambda}^{-1} &= \sum_{\lambda_{1}\lambda_{2}} \left\{ \frac{1}{2}\left(1+n_{\lambda_{1}}^{0}+n_{\lambda_{2}}^{0}\right)\zeta_{-} + \left(n_{\lambda_{1}}^{0}-n_{\lambda_{2}}^{0}\right)\zeta_{+} \right\},
\end{split}
\end{equation}	

\begin{equation}
\label{eq:FourP}
\begin{split}
\tau_{4,\lambda}^{-1} &= \sum_{\lambda_{1}\lambda_{2}\lambda_{3}} \left\{ 
\frac{1}{6} \frac{ n_{\lambda_1}^{0}n_{\lambda_2}^{0}n_{\lambda_3}^{0} }{ n_{\lambda}^{0} } \zeta_{--}
+\frac{1}{2} \frac{ \left(1+n_{\lambda_1}^{0}\right)n_{\lambda_2}^{0}n_{\lambda_3}^{0} }{ n_{\lambda}^{0} } \zeta_{+-}
+\frac{1}{2} \frac{ \left(1+n_{\lambda_1}^{0}\right)\left(1+n_{\lambda_2}^{0}\right)n_{\lambda_3}^{0} }{ n_{\lambda}^{0} } \zeta_{++}
   \right\},
\end{split}
\end{equation}	
with 
\begin{equation}
\label{eq:ThreeL}
\begin{split}
\zeta_{\pm} & = \frac{\pi\hbar}{4N} \vert V_{\pm}^{(3)} \vert^{2} \Delta_{\pm} \frac{ \delta ( \omega_{\lambda}\pm\omega_{\lambda_1}-\omega_{\lambda_2} ) }{ \omega_{\lambda}\omega_{\lambda_1}\omega_{\lambda_2} },
\end{split}
\end{equation}	

\begin{equation}
\label{eq:FourL}
\begin{split}
\zeta_{\pm\pm} & = \frac{\pi\hbar^2}{8N^2} \vert V_{\pm\pm}^{(4)} \vert^{2} \Delta_{\pm\pm} \frac{ \delta ( \omega_{\lambda}\pm\omega_{\lambda_1}\pm\omega_{\lambda_2}-\omega_{\lambda_3} ) }{ \omega_{\lambda}\omega_{\lambda_1}\omega_{\lambda_2}\omega_{\lambda_3}  },
\end{split}
\end{equation}	
where $V_{\pm}^{(3)}$ and $V_{\pm\pm}^{(4)}$ are in turn given by
\begin{equation}
\label{eq:ThreeV}
\begin{split}
V_{\pm}^{(3)} & = \sum_{ b,l_{1}b_{1},l_{2}b_{2} } \sum_{ \alpha\alpha_{1}\alpha_{2} } \Phi_{ 0b,l_{1}b_{1},l_{2}b_{2} }^{ \alpha\alpha_{1}\alpha_{2} } \frac{ e_{\alpha b}^{\lambda} e_{\alpha_1 b_1}^{\pm\lambda_1} e_{\alpha_2 b_2}^{-\lambda_2} }{ \sqrt{ m_{b} m_{b_1} m_{b_2} } } e^{\pm i \mathbf{k}_1 \cdot \mathbf{r}_{l_1} } e^{- i \mathbf{k}_2 \cdot \mathbf{r}_{l_2} },
\end{split}
\end{equation}	

\begin{equation}
\label{eq:FourV}
\begin{split}
V_{\pm\pm}^{(4)} & = \sum_{ b,l_{1}b_{1},l_{2}b_{2},l_{3}b_{3} } \sum_{ \alpha\alpha_{1}\alpha_{2}\alpha_{3} } \Phi_{ 0b,l_{1}b_{1},l_{2}b_{2},l_{3}b_{3} }^{ \alpha\alpha_{1}\alpha_{2}\alpha_{3} } \frac{ e_{\alpha b}^{\lambda} e_{\alpha_1 b_1}^{\pm\lambda_1} e_{\alpha_2 b_2}^{\pm\lambda_2} e_{\alpha_3 b_3}^{-\lambda_3} }{ \sqrt{ m_{b} m_{b_1} m_{b_2} m_{b_3} } } e^{\pm i \mathbf{k}_1 \cdot \mathbf{r}_{l_1} } e^{\pm i \mathbf{k}_2 \cdot \mathbf{r}_{l_2 } } e^{- i \mathbf{k}_3 \cdot \mathbf{r}_{l_3} }. 
\end{split}
\end{equation}	

In the above equations, $l$, $b$, and $\alpha$ are indexes of primitive cells, basis atoms, and cartesian coordinates, respectively; $m$ is the atomic mass, and $\mathbf{r}$ is the lattice vector of the primitive cell; 
$\mathbf{k}$, $n_{\lambda}^{0}$, $\omega_{\lambda}$, and $e^{\lambda}$ are phonon wave vector, equilibrium occupation number, frequency, and eigenvector, respectively; $\Phi_{ 0b,l_{1}b_{1},l_{2}b_{2} }^{ \alpha\alpha_{1}\alpha_{2} }$ and $\Phi_{ 0b,l_{1}b_{1},l_{2}b_{2},l_{3}b_{3} }^{ \alpha\alpha_{1}\alpha_{2}\alpha_{3} }$ are the third and fourth order IFCs, respectively.  In the three-phonon processes, $\zeta_{-}$ represents the  splitting process ($ \lambda \rightarrow \lambda_1 + \lambda_2 $) and $\zeta_{+}$ indicates the combination process ($ \lambda + \lambda_1 \rightarrow \lambda_2 $). Similarly, $\zeta_{--}$, $\zeta_{+-}$ and $\zeta_{++}$ represent the splitting ($ \lambda \rightarrow \lambda_1 + \lambda_2+ \lambda_3 $), redistribution ($ \lambda + \lambda_1 \rightarrow \lambda_2+ \lambda_3 $), and combination ($ \lambda + \lambda_1 + \lambda_2 \rightarrow  \lambda_3 $) in the four-phonon processes \cite{Tianli2016}. In both three- and four-phonon processes, momentum conservation is strictly enforced as indicated by $\Delta_{\pm}$ and $\Delta_{\pm\pm}$, and energy conservation is enforced by $\delta$ functions, which are approximated by adaptive and regular Gaussian smearing \cite{wuli2012,shengbte} in computing $\tau_{3,\lambda}^{-1}$ and $\tau_{4,\lambda}^{-1}$, respectively.

\subsubsection{Computational Details for BTE} 
In the BTE-ALD calculations, for both MgO and PbTe, we used 6$\times$6$\times$6 supercells to extract harmonic and anharmonic interatomic force constants (IFCs) up to the fourth order using compressive sensing lattice dynamics \cite{csld}. There is no explicit cutoff distance enforced on the harmonic IFCs \cite{Parlinski1997}. To further verify the extracted harmonic IFCs, we compare the calculated phonon dispersions with those independently obtained by Phonopy \cite{Phonopy}. The cutoff distance of the third order IFCs is limited to the seventh nearest neighbour shell, which leads to converged lattice thermal conductivity when only three-phonon interactions are accounted for. Considering the short-range nature of 4th-order anharmonicity \cite{csld} and the associated combinatorial explosion in the number of parameters, the 4th-order IFCs are limited to the second nearest neighbour shell. 
We also verified that the additional inclusion of the third nearest neighbour shell in the fourth order IFCs leads to negligible changes in both frequency shifts and lattice thermal conductivity.  The phonon BTE with renormalized harmonic IFCs and anharmonic IFCs as input were solved using $\mathbf{q}$-point mesh of 16$\times$16$\times$16 and 12$\times$12$\times$12, respectively, which are equivalent 
to the supercell structures used in MD simulations.

\subsubsection{Iterative and non-iterative solutions of BTE} \label{34scattering}

Phonon BTE under RTA can be linearized; when solved non-self-consistently,  their solutions are within the so-called single mode relaxation approximation (SMRTA) and when solved iteratively, they are within the relaxation time approximation (IRTA), respectively. Within the SMRTA, scattering rates arise from both normal ($N$) and Umklapp ($U$) processes, which lead to underestimation of $\kappa_{l}$ since $N$ processes do not introduce thermal resistance directly. For systems with significant $N$ processes, an iterative solution to the linearized BTE should be pursued by accounting for nonequilibrium phonon distribution through iteratively refining phonon populations \cite{OMINI1995101,omini2,Broido2005}. In this study, we calculate three-phonon scattering rates by solving the phonon BTE in an iterative manner. Considering the extremely high computational cost of including additional four-phonon scattering in the iterative solver, we treat four-phonon scattering non-iteratively, and   combine them with three-phonon scattering rates based on Matthiessen's rule to give the total scattering rates. We found that this strategy leads to reasonably accurate lattice thermal conductivity for compounds which are dominated by $U$ processes, such as PbTe, as discussed in detail in an earlier study \cite{YXPbTe2018}. It is worth noting that fully iterative soluation of BTE including both three- and four-phonon scatterings has been recently developed and implemented by Feng~\etal~\cite{Feng2018G} and Ravichandran~\etal~\cite{Ravichandran},respectively.   

\subsubsection{Temperature-dependent Frequencies: Anharmonic Phonon Renormalization} \label{prn}
 Several first-principles phonon renormalization (PRN) schemes based either on real or reciprocal space formalisms have been introduced to treat strong anharmonicity effects on phonon frequencies \cite{Souvatzis2009,Errea2014,Hellman2011,Tadano2015,Roekeghem2016}. In this work, we utilize our recently introduced real space based PRN scheme to compute temperature-dependent phonon freqeuncies and eigenvectors \cite{YXPbTe2018}. The temperature effects are taken into account via constructing temperature-dependent effective harmonic potential coefficients \cite{wallace1998thermodynamics} which includes the temperature-dependent corrections from higher-order IFCs on top of harmonic IFCs \cite{YXPbTe2018}. We refer the reader to Refs. [\onlinecite{YXPbTe2018,XiaGeTe2018}] for detailed discussions.

\subsection{Statistics}

\begin{figure}[htp]
    \centering
	\includegraphics[width = 0.6\linewidth]{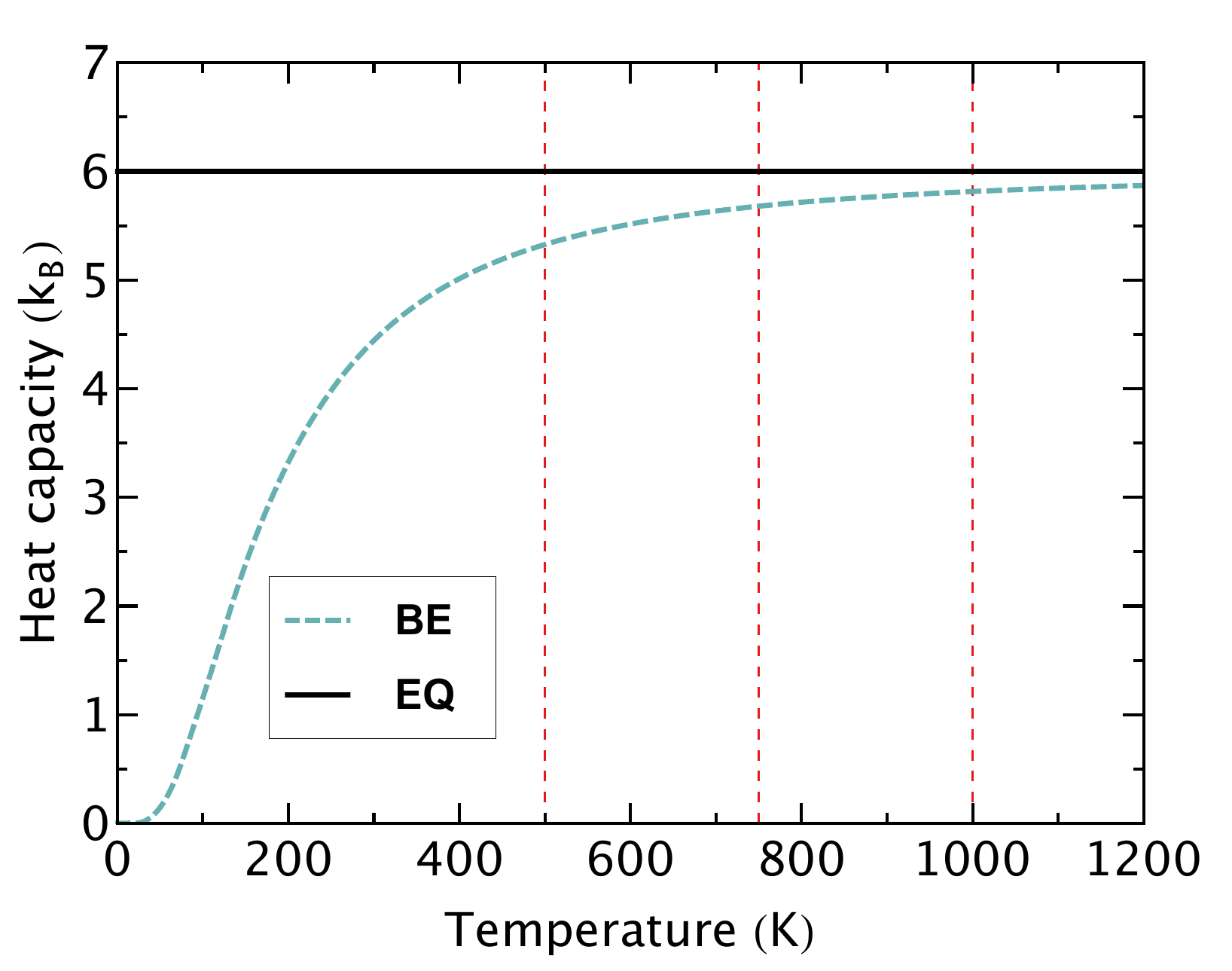}
	\caption{
        Constant volume heat capacity as a function of temperature for MgO. Blue dashed lines display the results computed using Bose-Einstein (BE) statistics and black solid lines denote  equipartition (EQ) as demonstrated in MD simulations. The dashed red lines indicate 500 K, 750 K and 1000 K.
	}.
	\label{fig:HC}
\end{figure}

In the expression of the thermal conductivity obtained from the BTE, the phonon statistics enter both the definition of the heat capacity and that of the scattering rates (inverse of relaxation time). According to Bose-Einstein (BE) statistics, the phonon population of mode $\lambda$ in equilibrium is
\begin{equation}
n_{\lambda}^{0} = \frac{1}{e^{\hbar\omega_{\lambda}/k_{\text{B}}T}-1},
\end{equation}
which leads to a mode heat capacity of 
\begin{equation}
C_{\lambda} = k_{\text{B}} \left( \frac{\hbar\omega_{\lambda}}{k_{\text{B}}T} \right)^2 \frac{e^{\hbar\omega_{\lambda}/k_{\text{B}}T}}{ \left( e^{\hbar\omega_{\lambda}/k_{\text{B}}T}-1 \right)^2}.
\label{phonon_BTE_Cv}
\end{equation}
In the classical (equipartition) limit, each phonon mode has an energy of $k_{\text{B}}T$ and population of $k_{\text{B}}T/\hbar\omega_{\lambda}$, thus giving rise to a temperature-independent mode heat capacity of $k_{\text{B}}$. The classical limit always overestimates the heat capacity (see Fig.\ \ref{fig:HC}), particularly at low temperatures. Specifically in the case of MgO, the classical limit leads to overestimates in the heat capacity of 11.2\%, 5.3\% and 3.1\% at 500 K, 750 K and 1000 K, respectively. 

As rigorously derived by Feng and Ruan \cite{Tianli2016}, phonon population in the classical limit (equipartition) cannot be directly used to compute three- and four-phonon scattering rates in Eq.\ \ref{eq:ThreeP} and \ref{eq:FourP}, since the resulting relaxation time is not properly defined from the linearized BTE. However, since classical limit and BE statistics do converge to the same limit at high temperature,
we utilized the equipartition phonon occupation in Eq.\ \ref{eq:ThreeP} and \ref{eq:FourP} to compute scattering rates in order to  compare with molecular dynamics simulations performed at high T.

\section{Results and Discussions}

\begin{figure}[h]
	\includegraphics[width = 1.0\linewidth]{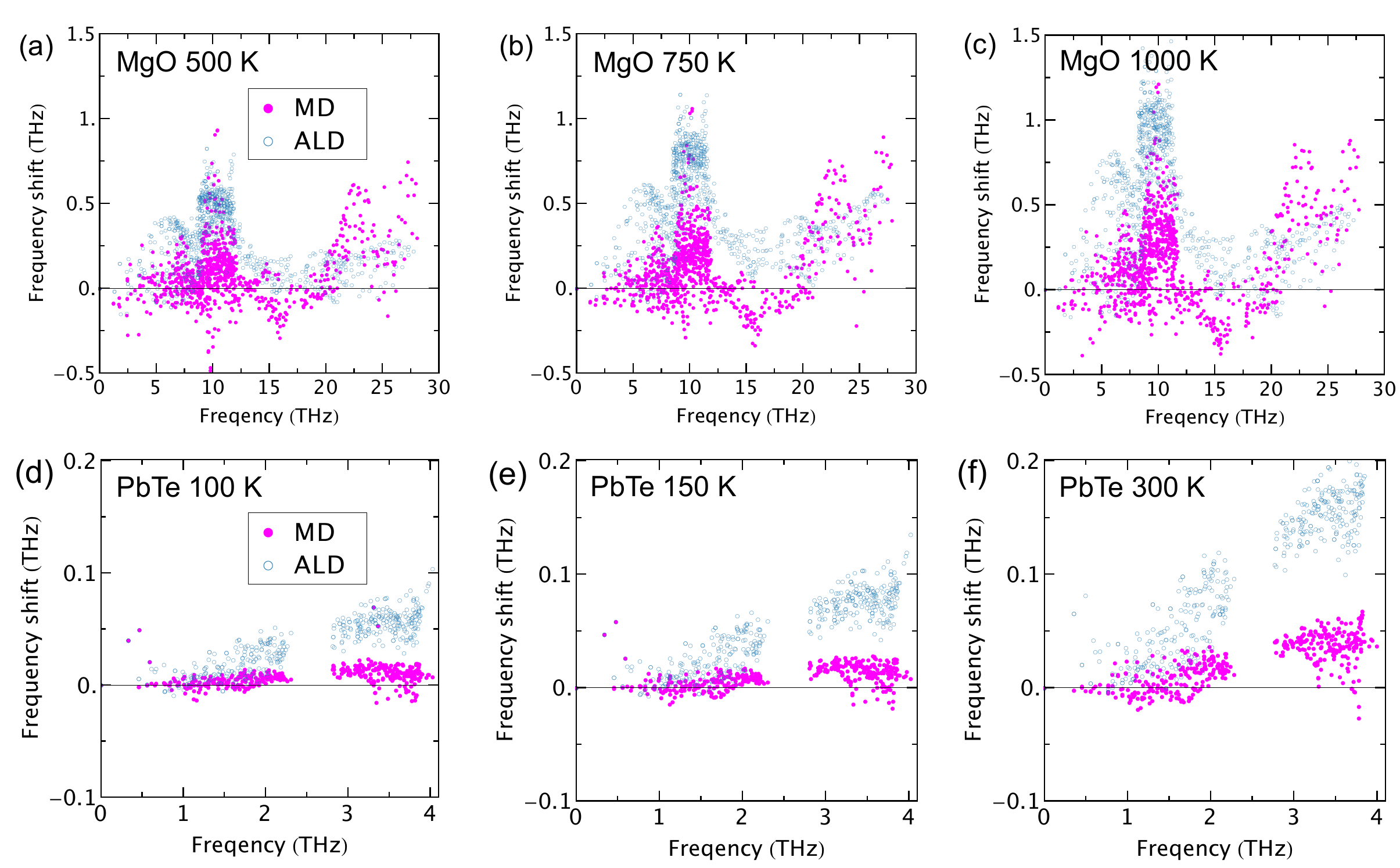}
	\caption{
        Mode-resolved frequency shift of MgO at finite temperatures [(a): 500 K; (b): 750 K; (c): 1000K] relative to the 0 K. Mode-resolved frequency shift of PbTe at finite temperatures [(d): 100 K; (e): 150 K; (f): 300K] relative to the 0 K. The solid magenta disks and empty blue circles denote results from molecular dynamics simulations and anharmonic phonon renormalization, respectively. 
        }.
	\label{fig:FreqShift}
\end{figure}

\subsection{Comparison of temperature-dependent phonon frequencies}
It is instructive to compare the phonon frequencies and their changes as a function of temperature obtained using the two different approaches. Fig. \ref{fig:dispersion} in the Supplemental Materials (SM)  shows comparisons of phonon dispersions, showing that the fitted potential using CSLD 
gives similar dispersions to those obtained using finite displacements with the original interatomic potential. 
In Fig. \ref{fig:FreqShift}, we show the mode-resolved frequency shifts as calculated in ALD and in MD, relative to the frequencies obtained from direct diagonalization. Fig. \ref{fig:DOS} in the SM displays the phonon density of states for these three cases. 
As expected, the frequency shifts increase with temperature. With the exception of high frequency MgO modes, the frequency shifts in MD are found to be smaller than the ones in ALD. The difference in frequency shifts between MD and ALD increases at high frequency, likely because of the different phonon distributions. This difference persists at high temperatures, despite an expected decrease in the difference between the two phonon distributions with increasing temperatures. Moreover, we note that the agreement between ALD and MD approaches is better in MgO, where frequency shifts are large ($\sim$1 THz), than in PbTe, where frequency shifts are on average 10 times smaller ($\sim$0.1 THz). To further confirm the ALD results, we performed additional self-consistent phonon (SCPH) calculations of frequency shifts in reciprocal space, as derived from many-body Green-function theory \cite{mahan2000many,Errea2011,Tadano2015}; the results are found to agree with those obtained by the real space-based PRN scheme. This indicates that part of the discrepancies between MD and ALD may be numerical in nature.

\subsection{Comparison of phonon lifetimes}
\begin{figure}[htp]
	\includegraphics[width = 1.0\linewidth]{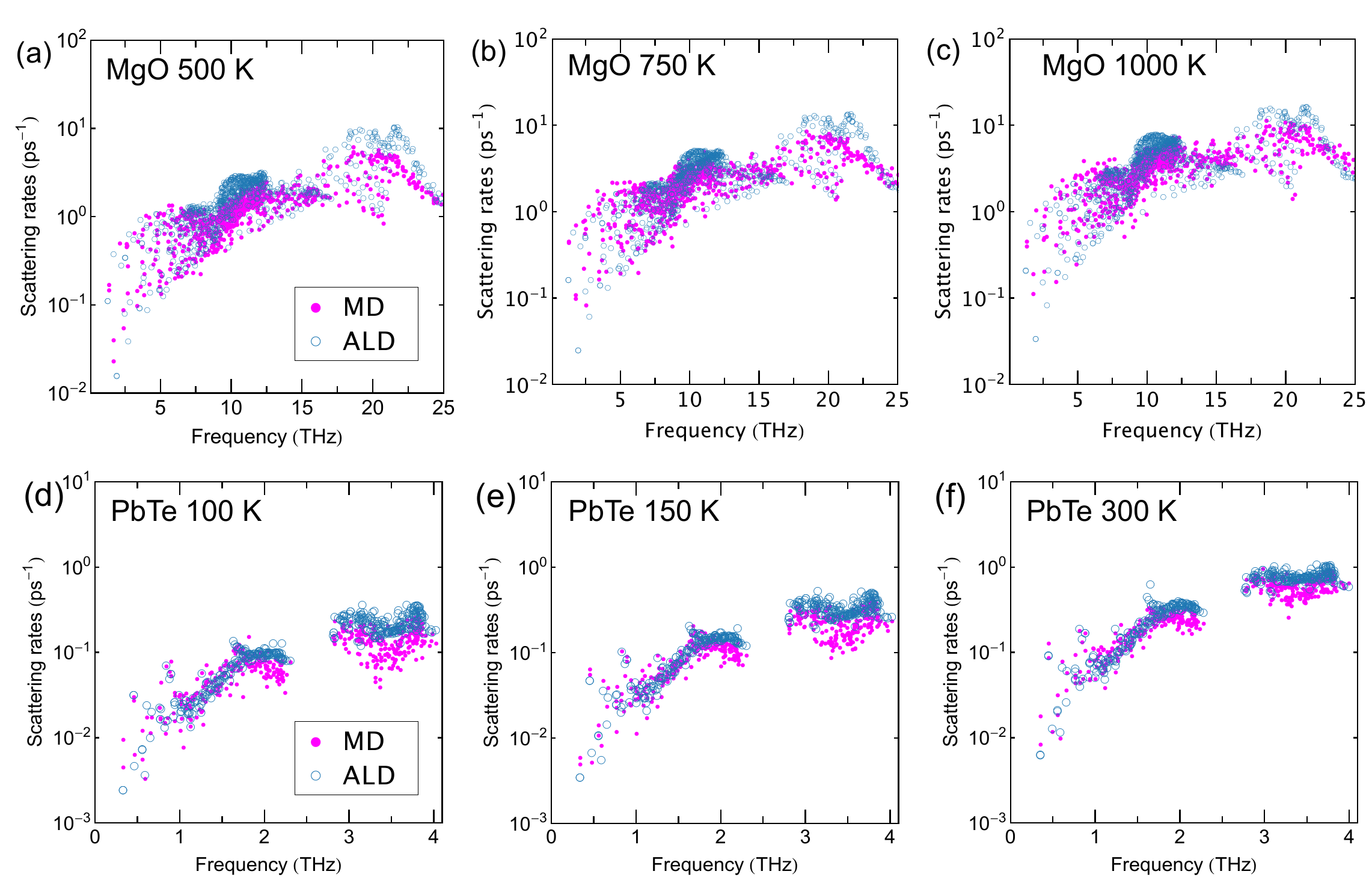}
	\caption{
        Comparison of mode-resolved scattering rates of MgO between molecular dynamics (MD) simulations and anharmonic lattice dynamics (ALD) calculations at (a) 500 K, (b) 750 K and (c) 1000 K. (d)-(f) the same as (a)-(c) but for PbTe at 100 K, 150 K and 300 K, respectively. Phonon populations were assumed to follow equipartition in ALD calculations.
	}.
	\label{fig:BTEMDrates}
\end{figure}

We compare the phonon mode-resolved scattering rates of MgO and PbTe obtained from MD and ALD at various temperatures in Fig. \ref{fig:BTEMDrates}. 
To perform a valid comparison, we enforce equipartition for phonon population in ALD calculations, thus following the statistics in MD simulations. Overall, good agreement is found between MD and ALD results, despite the rather different formalisms based on atomic motion or phonon quasiparticles. In general, acoustic modes show much smaller scattering rates than those of optical modes, primarily due to the limited scattering phase space of low-frequency phonon modes. 
Scattering rates of both acoustic and optical modes increase with enhanced phonon populations at higher temperatures. We notice that, for both MgO and PbTe, scattering rates of acoustic modes obtained from the two methods agree well with each other, while those of optical modes are found to be smaller from MD simulations, but with decreased discrepancy between MD and ALD at high temperatures. 
We note that the larger discrepancy associated with the high-lying optical modes at relatively low temperatures might be due to the break down of relaxation time approximation when equipartition is assumed for phonon population in the linearized BTE, while high temperature tends to reduce such discrepancy, as pointed out by by Feng and Ruan \cite{Tianli2016}. To shed light on the impact of statistics on lifetimes in ALD calculations, we also compare the scattering rates calculated using phonon populations obeying Bose-Einstein statistics and equipartition, respectively, in Fig. S5 (see Supplementary Materials). Consistently, we find that equipartition leads to higher phonon scattering rates, and again, with decreased difference from those obtained by assuming Bose-Einstein statistics at higher temperatures. This discrepancy is deeply rooted in the fact that phonon populations from Bose-Einstein statistics and equipartition are different, particularly for the high-frequency optical phonons. It is the overall increased phonon populations in the case of equipartition that leads to higher phonon scattering rates. Our results suggest that phonon scattering rates calculated using Bose-Einstein statistics compare better with MD simulations that those from equipartition, particularly for the high-lying optical phonon modes. This agreement, however, may be accidental or suggest error cancellations.
depend on the populations $n$ of various phonon modes. The effect of these differences is rather difficult to estimate, even limiting ourselves to three phonon processes in the SMRTA, due for example to the form of equation \ref{eq:ThreeP}, involving both a sum and a difference between different phonon populations.

\begin{figure}[htp]
	\includegraphics[width = 1.0\linewidth]{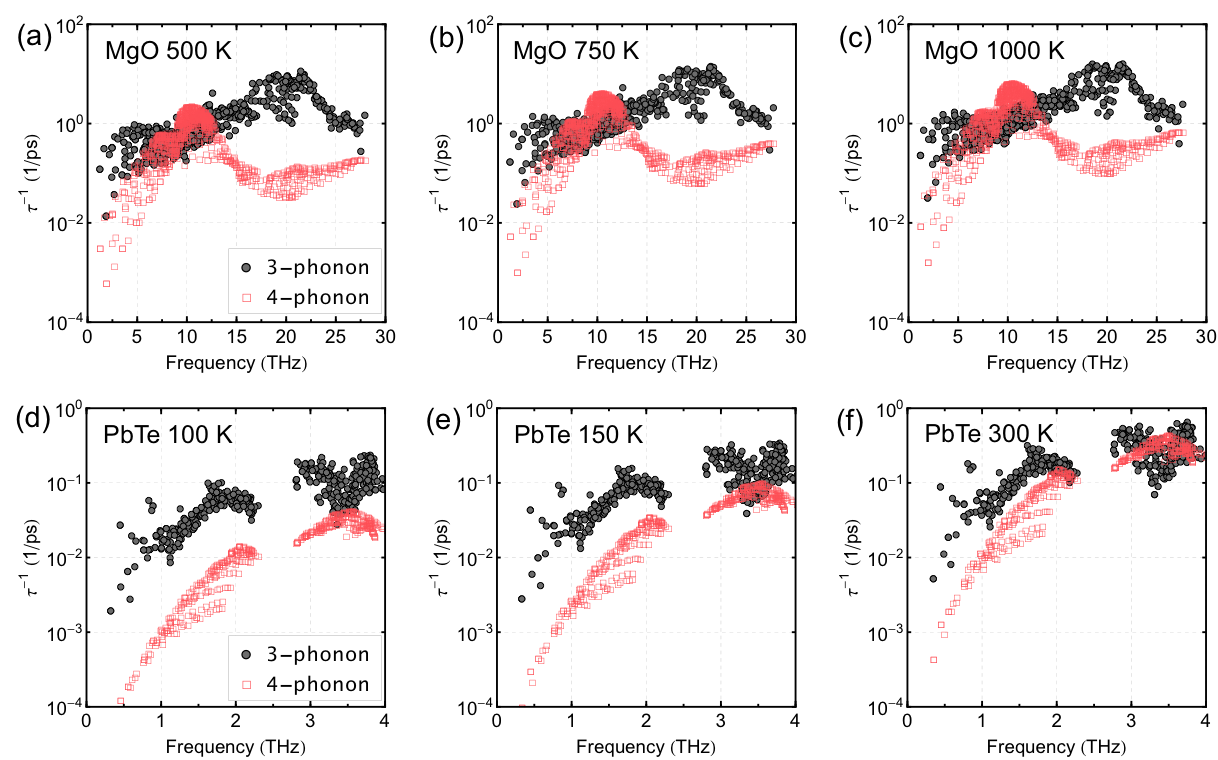}
	\caption{
	Decomposed three- and four-phonon scattering rates for MgO at (a) 500~K, (b) 750~K and (c) 1000~K, and PbTe at (d) 100~K, (e) 150~K and (f) 300~K.
	}.
	\label{fig:3and4rates}
\end{figure}

To reveal the role of the higher-than-third-order phonon scattering processes in determining overall phonon scattering rates, we show the decomposed total scattering rates by separating them into contributions from three- and four-phonon processes for MgO and PbTe in Fig. \ref{fig:3and4rates}. The MgO results reveal that the contribution of four-phonon scattering is comparable to that of  three-phonon scattering at all temperatures studied here. Particularly, four-phonon scattering rates of optical modes with frequency of about 10 THz are even higher than those of three-phonon processes, highlighting the importance of including four-phonon scattering processes to accurately predict the thermal conductivity. It is also evident from the PbTe results that four-phonon scattering rates have a stronger temperature-dependence than three-phonon scattering rates, with increasing relative contributions to total scattering rates at higher temperatures.

\subsection{Thermal Conductivity}

Here we discuss our finding, summarized in Table \ref{tab:kappa}, in which we use lifetimes obtained from MD simulations ($\tau_{MD}$), as well as from three- and four-phonon scattering ($\tau_{BE}$, in the BTE expression for thermal conductivity (Eq. \ref{eq:BTE}). We also use the classical equipartition ($C_{EQ}$) and Bose-Einstein ($C_{BE}$) expressions for heat capacity in Eq. \ref{eq:BTE}. All the above results are listed in comparison to the Green-Kubo (GK) thermal conductivity, and labeled approximations 1-7 or A1-A7 in Table \ref{tab:kappa}. For MgO, the thermal conductivity values obtained from BTE using three- and four-phonon scattering processes, and Bose-Einstein statistics (A4 in Table \ref{tab:kappa}), agrees strikingly well (to within 3\%) with the Green-Kubo results (A7). For PbTe, however, the agreement is only satisfactory (within 16\%). This agreement is better than previous comparisons (e.g. Ref. \cite{Turney2009a}), likely due to the inclusion of fourth-order phonon scattering in the BTE-ALD treatment. The agreement, however, may be due to a number of error cancellations, as explained below.\\

\begin{table}[h]
\begin{tabular}{| l | l | l | l | l | l | l | l | l | }
 \hline
& & A1 & A2 & A3 & A4 & A5 & A6 & A7  \\
 \hline
 Compound & Temperature & $\tau_{MD}$ $C_{EQ}$ & $\tau_{MD}$ $C_{BE}$ & $\tau_{BE}$ $C_{EQ}$ & $\tau_{BE}$ $C_{BE}$ & $\tau_{EQ}$ $C_{EQ}$ & $\tau_{EQ}$ $C_{BE}$ & GK           \\
\hline
 MgO & 500K        & 29.4             & 27.0         & 35.5     & 32.8    & 27.1   & 25.2   & 32.6 (2.1)  \\
 MgO & 750K        & 16.3             & 15.6         & 20.9     & 20.2    & 17.4   & 16.8   & 19.6 (0.5)  \\
 MgO & 1000K       & 10.4             & 10.2         & 14.7     & 14.4    & 12.6   & 12.4   & 14.1 (0.7)  \\
\hline
 PbTe & 100K        &11.2             & 10.7         & 13.0     & 12.4    & 10.2   & 9.8   & 10.4 (0.8)   \\
 PbTe & 150K        & 7.0             & 6.8         & 7.9     & 7.7    & 6.6   & 6.5   & 6.5 (0.6)   \\
 PbTe & 300K        & 3.1             & 3.1         & 3.4     & 3.4    & 3.0   & 3.0   & 3.2 (0.2)  \\
\hline
  \end{tabular}
\caption{Computed thermal conductivity of MgO and PbTe using different expressions for lifetimes ($\tau$) and heat capacities ($C$). All thermal conductivity values are in $\text{W/mK}$. The lifetimes $\tau_{MD}$'s are obtained from MD simulations according to Equation \ref{eq:MD_tau}, $\tau_{BE}$'s are obtained from BTE according to phonons in a Bose Einstein distribution, whereas $\tau_{EQ}$'s are obtained from BTE according to phonon population obeying equipartition.  All BTE calculations include three- and four-phonon processes. The heat capacities $C_{EQ}$ and $C_{BE}$ correspond to those obtained from equipartition and Bose-Einstein statitics, respectively, as shown in Fig. \ref{fig:HC}. Finally, the thermal conductivity obtained from Green-Kubo (GK) is listed. The different approximations are labeled A1 to A7 in the first row. }
\label{tab:kappa}
\end{table}

One significant difference between the MD and BTE approaches is the inclusion of phonon scattering processes to all orders in MD, as opposed to the order-by-order expansion in BTE. Because the rate of scattering processes at different orders are summed, i.e. via Matthiessen's rule, the inclusion of higher order should reduce the lifetimes and therefore the thermal conductivity. Indeed, substituting lifetimes from MD into the BTE expression with the BE expression for heat capacity (A2) results in lower thermal conductivity values compared to using BTE lifetimes (A4), due to the inclusion of higher-order scattering processes in MD. Because of the computation of the lifetimes using energy autocorrelation functions, the dissipative vs non-dissipative effects of umklapp vs normal processes, respectively, are also preserved.

The effects of statistics on the heat capacity and the corresponding effects on thermal conductivity can be made apparent by considering the changes in predicted values as the heat capacity is evaluated using equipartition (A1, A3, A5) or Bose-Einstein statistics (A2, A4, A6), when the treatment of the lifetimes is held constant. Since classical heat capacities are consistently higher, and especially so at low temperatures, than Bose-Einstein heat capacities, the thermal conductivity values are also higher in A1, A3, and A5 than in A2, A4, and A6, respectively. The difference decreases with increasing temperature as expected. Moreover, we find that using BE heat capacities with MD lifetimes (A2) worsens rather than improves the agreement with the BTE-ALD results (A4). In agreement with Ref. \cite{Turney2009b}, BE treatment of heat capacities together with MD lifetimes is ruled out as a possible quantum correction for classical MD simulations of thermal conductivity. 

As mentioned above, phonon occupations also directly affect scattering rates calculated from three- and four-phonon scattering processes. This effect is apparent from a comparison between A4 and A6. It is seen that changing the occupation numbers to those corresponding to equipartition (A6) rather than the Bose-Einstein distribution (A4), without changing the heat capacity, changes the thermal conductivity significantly (up to 24\%). Note also, that the effect of statistics on scattering rates is significantly larger than the effects of statistics on heat capacities. For example, the differences between A5 and A6 is smaller than those between A4 and A6. 

Interestingly, substituting MD lifetimes as well as heat capacities evaluated with equipartition into the BTE expression (A1) results in thermal conductivity values which are significantly different from those obtained from the Green Kubo expression (A7). The difference is likely due to the single-mode approximation used in the BTE approach. This highlights the fact that apart from treatments of lifetimes and heat capacities, the summation according to phonon modes in and of itself introduces a difference between BTE and MD methods for computing thermal conductivity. 

\section{Conclusion}
In conclusion, we have presented a detailed comparison between molecular dynamics (MD) and Boltzmann Transport Equation using anharmonic lattice dynamics (BTE-ALD) approaches for the computation of thermal conductivity. Improvements in BTE-ALD such as the inclusion of fourth-order phonon scattering processes, and treatment of temperature-dependent phonon frequency shifts (phonon renormalization) were included. Issues regarding the proper treatment of statistics in MD simulations, namely classical equipartition rather than Maxwell Boltzmann, were addressed. Thermal conductivity values were found to agree well between BTE-ALD and Green-Kubo (GK) MD, but a detailed analysis showed that such agreements are introduced by cancellations of different errors. By substituting lifetimes derived from energy-energy autocorrelation function from MD simulations into the BTE expression, we determined the effects of higher-than-4th-order phonon processes, the effects of statistics via the lifetimes and heat capacities, and the effects of single-mode relaxation time approximation itself, to the calculation of thermal conductivity. Significantly, we find that the effects of statistics on thermal conductivity is primarily due to effects on scattering rates. We also find that a full substitution of MD lifetimes and heat capacities in the BTE expression fails to reproduce GK results, indicating a significant effect of single-mode relaxation time treatment on thermal conductivity. One significant lesson from this comparison is that for integrated properties such as transport coefficients, apparent agreement may mask fundamental physical differences, thus caution is advised in the interpretation of the results.

\section*{Acknowledgment}
This work was supported by the Midwest Integrated Center for Computational Materials (MICCoM) as part of the Computational Materials Sciences Program funded by the U.S. Department of Energy, Office of Science, Basic Energy Sciences, Materials Sciences and Engineering Division (5J-30161-0010A). Use of the Center for Nanoscale Materials, an Office of Science user facility, was supported by the U.S. Department of Energy, Office of Science, Office of Basic Energy Sciences, under Contract No. DE-AC02-06CH11357. This research used resources of the National Energy Research Scientific Computing Center, a DOE Office of Science User Facility supported by the Office of Science of the U.S. Department of Energy under Contract No. DE-AC02-05CH11231.

\bibliography{MDBTE}

\widetext
\clearpage
\newpage
\begin{center}
	\textbf{\large Supplemental Materials: Computational prediction of lattice thermal conductivity -- a comparison of molecular dynamics and Boltzmann transport approaches}
\end{center}

\setcounter{equation}{0}
\setcounter{figure}{0}
\setcounter{table}{0}
\setcounter{section}{0}
\setcounter{page}{1}
\makeatletter
\renewcommand{\thepage}{S\arabic{page}}
\renewcommand{\theequation}{S\arabic{equation}}
\renewcommand{\thefigure}{S\arabic{figure}}
\renewcommand{\thetable}{S\arabic{table}}
\renewcommand{\bibnumfmt}[1]{[S#1]}
\renewcommand{\citenumfont}[1]{S#1}



\section{Interatomic force constants from Compressive sensing lattice dynamics}  \label{csld}

\begin{figure}[htp]
    \centering
	\includegraphics[width = 0.75\linewidth]{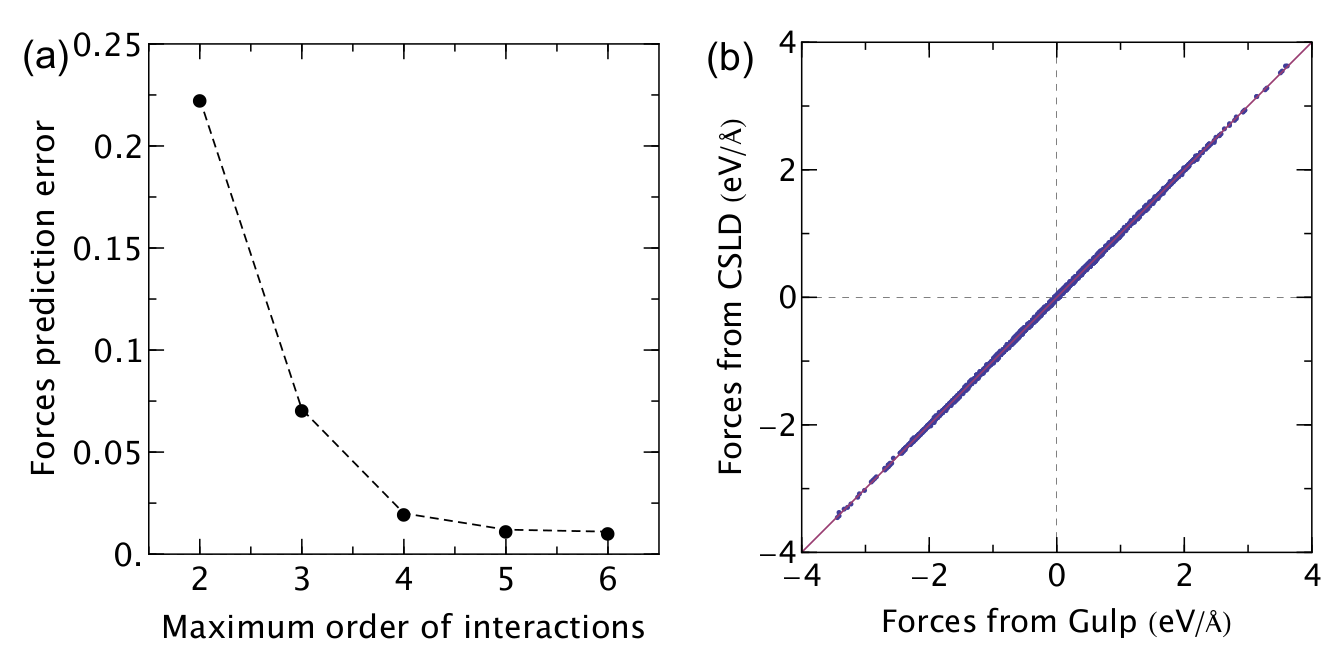}
	\caption{
        (a) Relative force prediction error by compressive sensing lattice dyanmics as a function of included maximum order of interactions. (b) Comparison between predicted and Gulp-computed forces when interatomic force constants up to 6th-order are included.
	}.
	\label{fig:IFC}
\end{figure}
The essential ingredients for the calculation of $\tau_{\lambda}^{-1}$ are harmonic and anharmonic interatomic force constants (IFCs). We utilized our recently-developed compressive sensing lattice dynamics (CSLD) \cite{csld} approach to extract both harmonic and anharmonic IFCs, which are obtained from a convex optimization problem that minimizes a weighted sum of the $\ell_{1}$ norm of the IFCs $\mathbf{\Phi}$ and the root-mean-square fitting error
	\begin{equation}
	\begin{split}
	\mathbf{\Phi}^\text{CS} &= {\arg \min}_\mathbf{\Phi} \; \| \mathbf{\Phi} \|_1 + \frac{\mu}{2} \| \mathbf{F} -  \mathbb{A} \mathbf{ \Phi} \|^2_2, 
	\end{split}
	\label{eq:L1min}
	\end{equation}
where $\mathbf{F}$ is a vector composed of atomic forces calculated by DFT, and $\mathbb{A}$ is a matrix formed by the products the atomic displacements. The parameter $\mu$ is a tuning parameter used to control the relative weights of the force fitting error versus the sparsity of $\mathbf{\Phi}$ components. Since (1) small $\mu$ leads to sparse solution of $\mathbf{\Phi}$ at the expense of the accuracy of the fitted $\mathbf{\Phi}$ (``underfitting'') and (2) large $\mu$ give a dense solution of $\mathbf{\Phi}$ but with poor predictive power (``overfitting''), an optimal $\mu$ is determined by monitoring the predictive relative error for a leave-out subset of the training data not used in fitting. To reduce the parameter space of $\mathbf{\Phi}$, linear constraints from crystal symmetry and translational symmetry are explored and strictly enforced by finding the null space representation of $\mathbf{\Phi}$. More details on numerical issues and the  symmetrization of IFCs can be found in Ref.~[\onlinecite{csld,vo2yi,csce1,csce2,Tadano2015}]. The dependence of force prediction error on the order of included interactions is given in \ref{fig:IFC}

\section{Phonon Dispersion and Density of States}
\begin{figure}[h]
	\includegraphics[width = 1.0\linewidth]{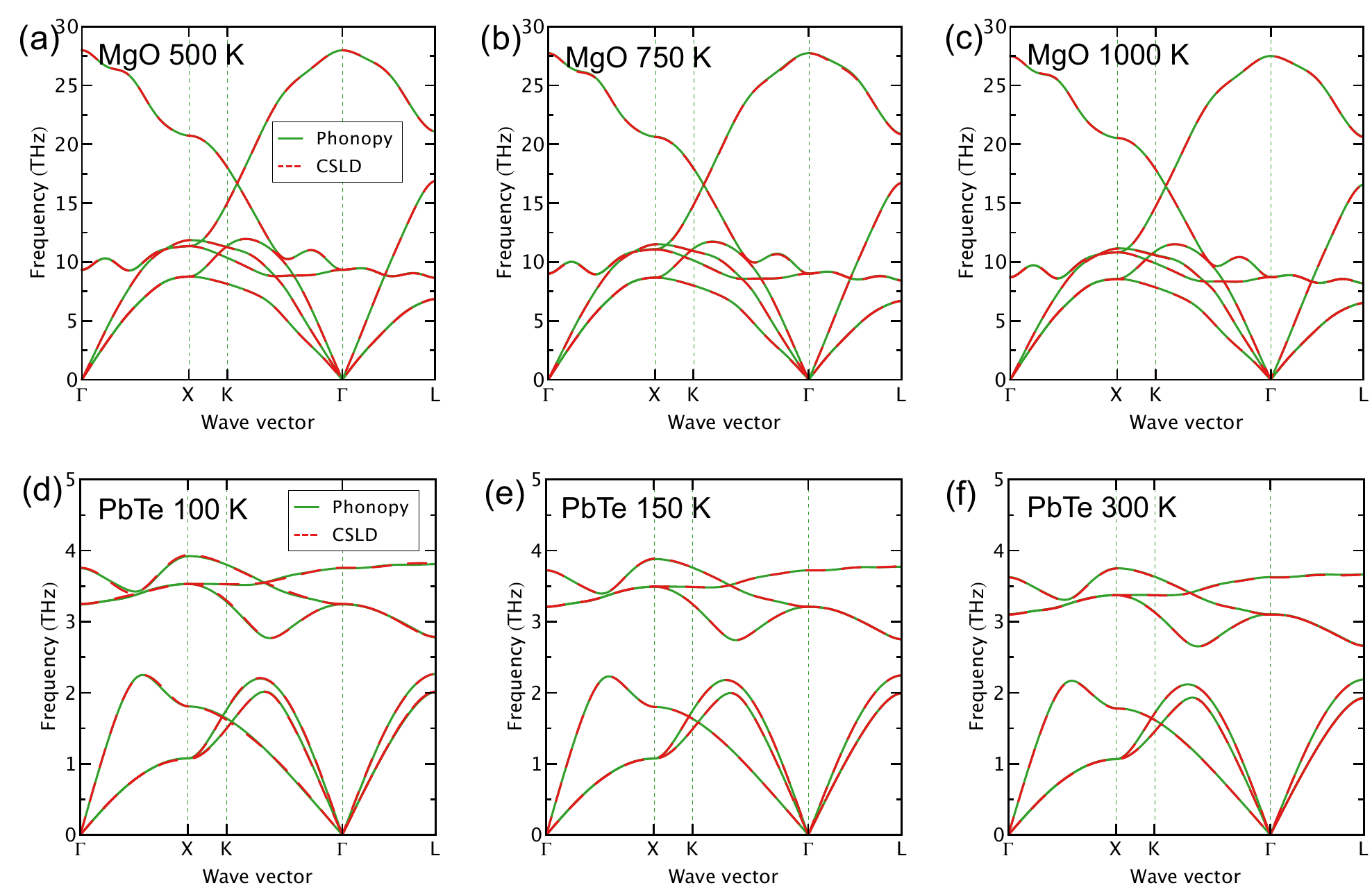}
	\caption{
        Comparison of phonon dispersions for MgO and PbTe calculated using Phonopy and compressive sensing lattice dyanmics (CSLD). Temperature-dependent lattice parameters were determined by MD simulations.
        	}.
	\label{fig:dispersion}
\end{figure}

\begin{figure}[h]
	\includegraphics[width = 1.0\linewidth]{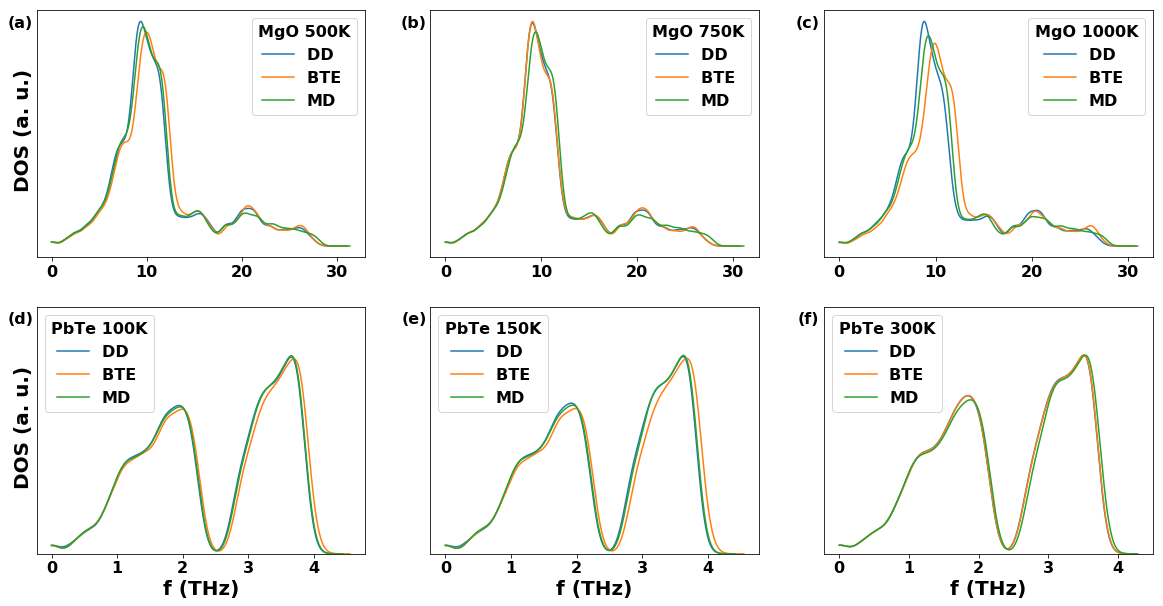}
	\caption{
	Phonon density of states plots for MgO at 500K (a), 750K (b), 1000K (c) and for PbTe at 100K (d), 150K (e) and 300K (f). DD (blue line) indicates the density of state calculated from the direct diagonalization of the dynamical matrix; BTE (orange) is the result of anharmonic lattice dynamics calculations and MD (green) is the result of molecular dynamics simulations.}
	\label{fig:DOS}
\end{figure}

\begin{figure}[h]
	\includegraphics[width = 1.0\linewidth]{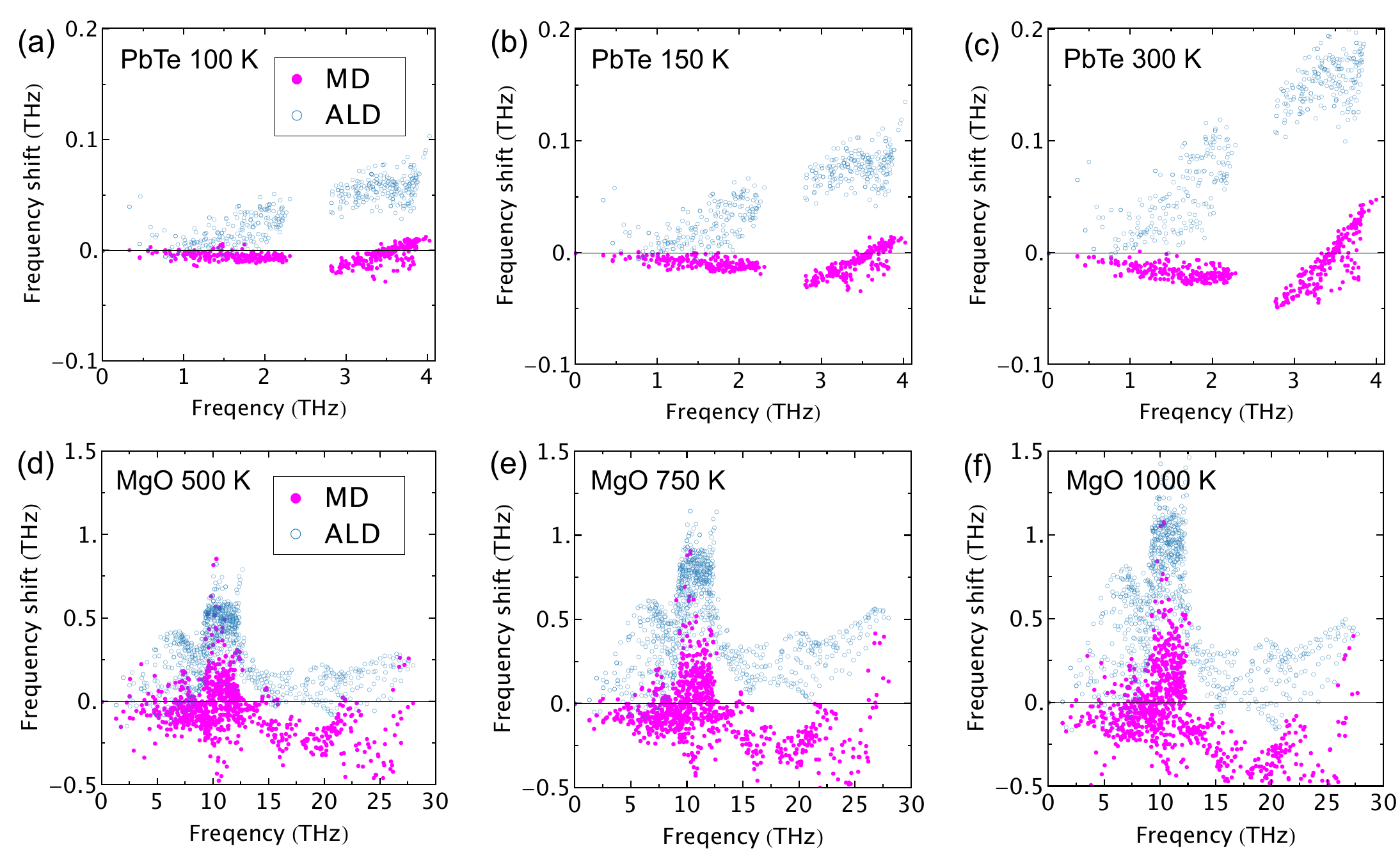}
	\caption{
        Mode-resolved frequency shift of MgO at finite temperatures [(a): 500 K; (b): 750 K; (c): 1000K] relative to the 0 K. Mode-resolved frequency shift of PbTe at finite temperatures [(d): 100 K; (e): 150 K; (f): 300K] relative to the 0 K. The solid magenta disks and empty blue circles denote results from molecular dynamics simulations and anharmonic phonon renormalization, respectively. Different from Fig.~4 in the main text, here the phonon frequency from molecular dynamics is obtained combining Eq \ref{eq:UK} and equipartition by taking the ratio $\| q_{\lambda} \|^2 / \| \dot{q}_{\lambda} \|^2$
        	}.
	\label{fig:FreqShiftNew}
\end{figure}

\begin{figure}[h]
	\includegraphics[width = 1.0\linewidth]{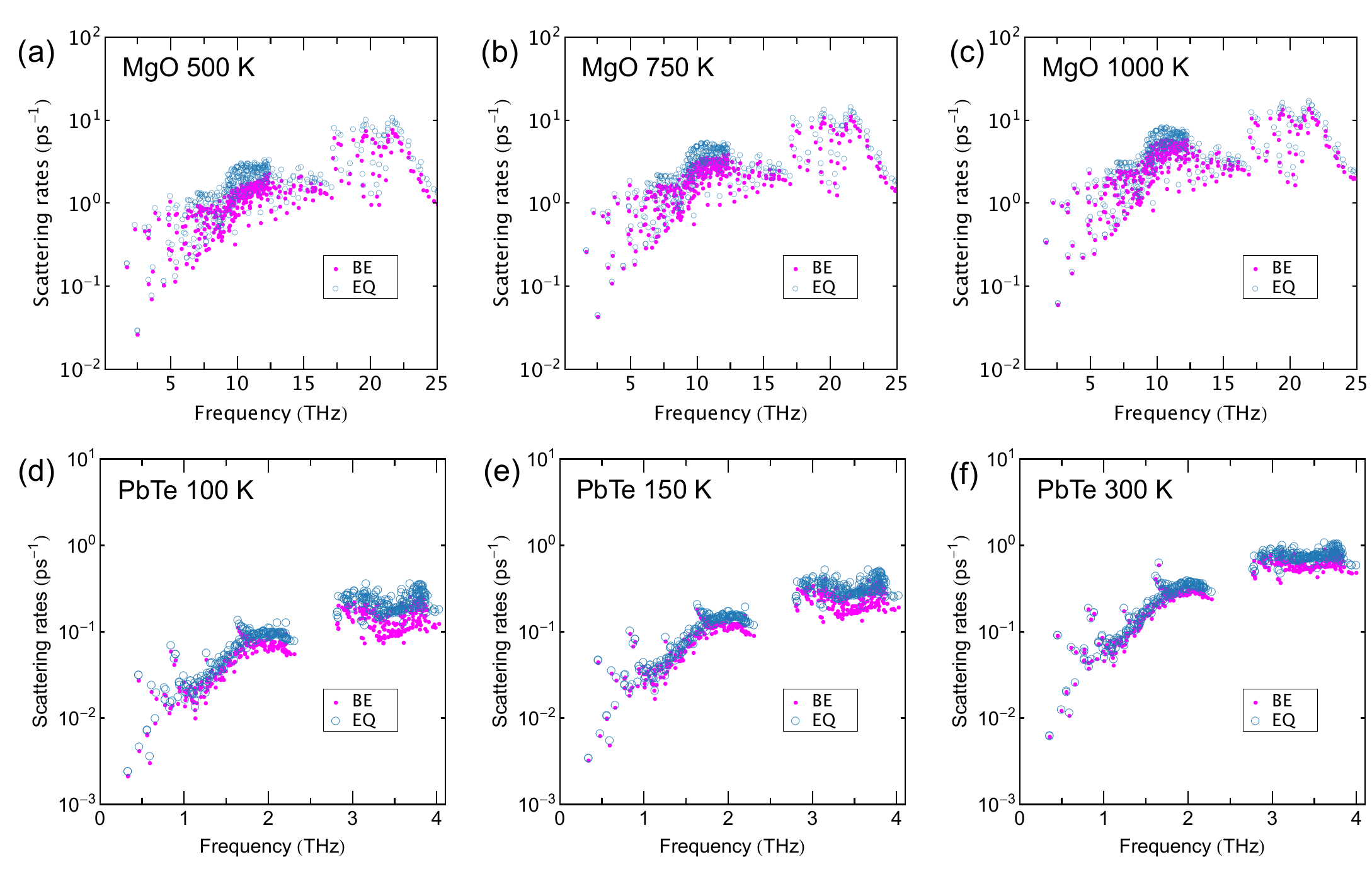}
	\caption{
        Comparison of mode-resolved scattering rates of MgO obtained by assuming Bose-Einstein statistic (BE) and classical equipartition (EQ) in anharmonic lattice dynamics (ALD) calculations at (a) 500 K, (b) 750 K and (c) 1000 K. (d)-(f) the same as (a)-(c) but for PbTe at 100 K, 150 K and 300 K, respectively.
        	}.
	\label{fig:lifetimeEQBE}
\end{figure}

\begin{figure}[h]
	\includegraphics[width = 1.0\linewidth]{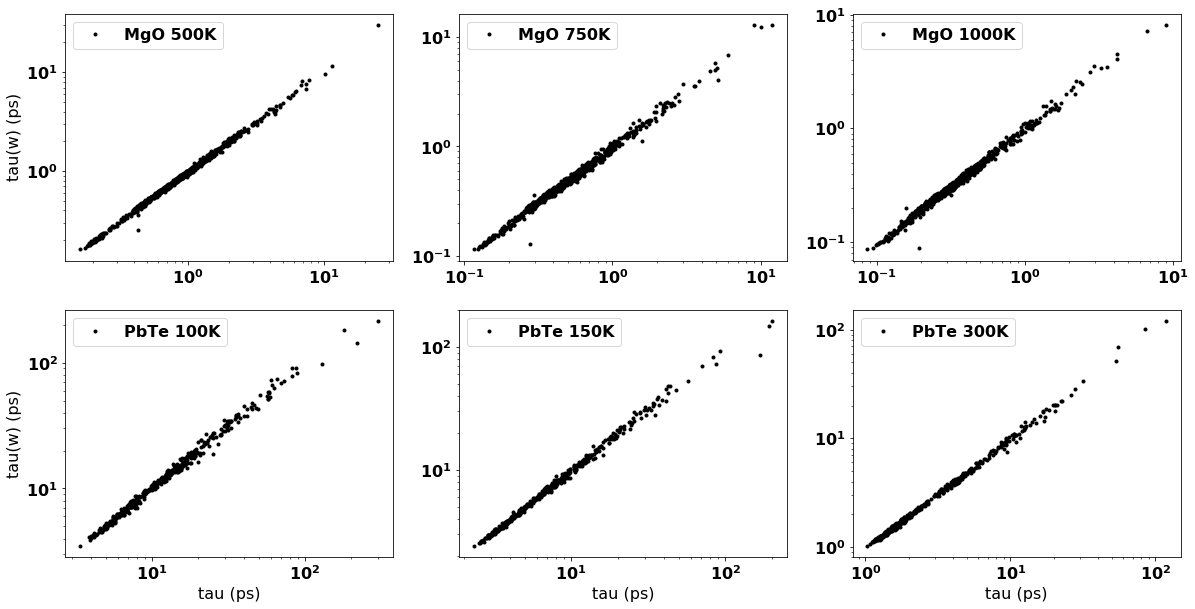}
	\caption{Lifetimes computed from Molecular Dynamics. The lifetimes calculated from Eq. \ref{eq:MD_tau} and Eq. \ref{eq:lifetime_cos_exp} are on the abscissa and ordinate, respectively.
	}
	\label{fig:DOS}
\end{figure}

\clearpage

\bibliography{MDBTE}

\end{document}